# THE EFFECT OF A SOUTHWARD INTERPLANETARY MAGNETIC FIELD ON STÖRMER'S ALLOWED REGIONS

J.F. Lemaire

*UCL-ASTR, Chemin du Cyclotron, B-1348 Louvain-la-Neuve*

*IASB, 3 Ave Circulaire, b-1180 Bruxelles, Belgium*

**ABSTRACT.** The motion of a charged particle in a magnetic dipole has first been studied by Störmer. The different applications of Störmer's theory to aurorae, cosmic rays and Van Allen radiation belt particles are recalled in an historical perspective. In this paper, we expand the Störmer theory in order to take into account the effects produced by an additional uniform and stationary interplanetary magnetic field (IMF) whose orientation is parallel or antiparallel to the magnetic moment of the dipole. A new expression is derived for the Störmer potential taking into account the additional IMF component. It is shown how Störmer's allowed and forbidden zones are influenced by the implementation of a northward or a southward IMF, and how a southward turning of the IMF orientation makes it easier for Solar Energetic Particle and Galactic Cosmic Rays to enter into the inner part of the geomagnetic field along interconnected magnetic field lines. [*]

## HISTORY AND APPLICATIONS OF STÖRMER'S THEORY

In 1907, the Norwegian mathematician Carl Störmer calculated numerically different classes of trajectories of charged particles moving in a dipole magnetic field like that of Birkeland's Terrella experiment. The aim of his study was to determine how electrons (cathode rays) could travel from the Sun, and penetrate across the geomagnetic field to produce eventually the aurorae borealis in the upper atmosphere of the Earth.

Störmer showed that in addition to the total energy of a particle moving in a dipole magnetic field, the azimuthal component of its generalized momentum, $p_\phi(r, t)$, is also a constant of motion. The conservation of these two physical quantities stems from (i) the assumed time independence of the magnetic field, $\boldsymbol{B}(r)$, and vector potential $\boldsymbol{A}(r)$ [$\boldsymbol{B} = \mathrm{curl}\,\boldsymbol{A}$], and from (ii) their cylindrical symmetry with respect to the axis of the magnetic dipole: i.e. their independence with respect to the azimuthal coordinate, $\phi$ (or $\phi$-invariance).

When the magnetic vector potential, $\boldsymbol{A}$, is stationary and the electric forces neglected, as in Störmer's theory, the Hamiltonian of a relativistic particle has a simple form that resembles a non-relativistic Hamiltonian

$$H(r,p) = (p - Ze\,A)^2 / 2m \tag{1}$$

Since the Hamiltonian is independent of time, the energy is a constant of motion as well as the velocity, $v$; the relativistic mass is defined by $m = m_o / (1 - v^2/c^2)^{1/2}$, where $m_o$ is the rest mass of the charged particle, and $Ze$ its electric charge.

---

The trajectories of particles are confined within Störmer's allowed (or permitted) regions. The meridional cross-section of these regions is illustrated by the white zones of Figure 1. This figure is similar to a well referenced and famous photograph published in Störmer's 1955 seminal Monograph. The solid line contours are the cross sections of the surfaces separating forbidden and allowed/permitted zones for different values of $\gamma$, a dimensionless constant which is proportional to the value $p_\phi$, and inversely proportional to the Earth's magnetic dipole moment : $M = 8.06 \; 10^{15}$ Tesla m$^3$. The value of $\gamma$ can assume any value between $\pm \infty$, like the value of $p_\phi$, the azimuthal component of the generalized momentum of the charged particle.

For $\gamma < -1$ the inner allowed region is closed; it contains permanently trapped particles that oscillate around a curve called "Thalweg". The Thalweg is shown by the dotted lines in the panels of Stormer's figure which is the couneterpart of our Figure 1. For particles of small energies $\gamma << -1$, and the Alfvén conditions are then satisfied. Under these conditions the gyroradius is small compared to the characteristic scale length of the non-uniform magnetic field; the standard guiding center approximation is then applicable to describe the motion of charged particles.
In this limit (i.e. for large negative values of $\gamma$) the guiding center of the particle oscillates between two conjugate mirror points located along the Thalweg. The Thalweg coincides precisely with a dipole field line. The guiding center moves approximatively along this magnetic field line, at least when there are no disturbing electric fields, or when their effects can be neglected in the first order approximation. This is generally assumed to be the case for charged particles with energies larger than a few hundred keV, which form the Van Allen belts.

The aurorae borealis are produced by electrons of less than 5 keV for which $\gamma$ is much smaller than −1. These particles are either confined to the closed inner allowed zone or to the outer open one; but there is no way for auroral electrons nor solar electrons of that relatively small energy and azimuthal component of the generalized momentum to traverse the forbidden region (in black) separating the two allowed ones (white areas). This forbidden region is a magnetic potential barrier that only particles with much higher energies are able to overcome, provided the azimuthal component of their generalized momentum $p_\phi$ is such that $-1 < \gamma < 0$. Therefore, Störmer's theory did not meet with success in explaining the expected solar origin of auroral electrons. Furthermore, note that for auroral particles whose energies are significantly less than 100 keV the effects of magnetospheric electric potentials of the order of 40-50 kV cannot be neglected. As a matter of consequence, Störmer's theory ignoring the presence of magnetospheric electric fields, is not applicable to auroral electrons and protons.

But two decades later, in the 30's, Störmer's theory proved to be much more adequate to account for the geomagnetic cut-off of the much more energetic Galactic Cosmic Rays (GCR). Indeed, these particles have energies exceeding 1 GeV, i.e. much larger than the above mentioned critical value of "a few hundred keV". Using the Störmer theory it has then been possible to explain the latitudinal dependence of the flux of Cosmic Rays. Galactic Cosmic Ray protons coming from infinity have a large enough magnetic rigidity ( $mv/Ze$ ) to overcome the magnetic potential barrier: they can penetrate into the geomagnetic field within the open allowed regions illustrated in Figure 1 for $-1 < \gamma < 0$.

Furthermore, when $-1 < \gamma < 0$, primary cosmic rays may hit the Earth's atmosphere at geomagnetic latitudes larger than the so-called "latitude of geomagnetic cut-off". Using Liouville's theorem and assuming an isotropic flux for these relativistic particles at infinity, Lemaître and Vallarta (1933) determined the distribution of the angles of incidence of Cosmic Ray particles in the upper atmosphere. This led them to account correctly not only for the latitudinal effect, but also for the observed east-west asymmetry of Cosmic Ray fluxes.

In addition to Störmer's formulation, other mathematical variables and coordinates have been proposed and used to approximate the motion of a charged particle in a dipole magnetic field. The most remembered contributions are Graef and Kusaka (1938), De Vogelaere (1954), Gall (1963),

Garmire (1963), Hamlin et al. (1961), Hayakawa and Obayashi (1963), Hones (1963), Ray (1963), Dragt (1965).

After the discovery of the Earth's Radiation Belts in 1958, the theory of Störmer was once again restored and popularized. It was then successfully applied and expanded to study the trajectories of 0.1-10 MeV electrons and 0.5-300 MeV protons magnetically trapped along the (nearly dipole) geomagnetic field lines. These trapped particles are confined to closed annular zones corresponding to Störmer's inner allowed region for $\gamma << -1$. The ensemble of allowed regions forms the Van Allen Radiation Belts, as well as the Ring Current of magnetospheric ions and electrons, the latter having kinetic energies in the range 10-100 keV.

As already mentioned, the guiding center approximation introduced by Alfvén (1940, 1950) can be used to describe the motion of such charged particles when their gyroradius is much smaller than the scale length of the magnetic field. Bossy (1962, 1963) showed how Alfvén's guiding center approximation can be derived from Störmer's theory as a first order approximation when the Larmor gyroradius is small compared to the characteristic scale length of the B-field. Singer (1958b) applied this principle in discussing the spatial extent of the inner Radiation Belt protons. The work of these pioneers was comprehensively reviewed by Morfill and Scholer (1973), Williams (1987).

Before the discovery of the Van Allen belts in 1958, the closed inner region of Störmer was assumed to be empty (Van Allen *et al.*, 1958, 1959). Indeed, before that date the physical mechanisms populating the inner allowed zone were not yet discovered. Note, however, that Singer (1957) had proposed, already a year earlier, to form a ring current of charged particles produced by nuclear interactions of primary cosmic ray with the atoms of the atmosphere. Although these secondary ions contribute, to form a ring current which circulates within this closed annular region, just above the atmosphere, unfortunately, the charged particles drifting at these low altitudes have a short life time: much too short to account for the extremely high fluxes observed later on, in outer space, with the first artificial satellites. It was only after the experimental discovery of the Van Allen belts that Singer (1958b) and Vernov *et al.* (1958, 1959a & b) independently proposed to fill up the inner allowed zone of Störmer with energetic protons originating from the beta decay of neutrons produced by nuclear interactions of primary Galactic Cosmic Rays impacting on atmospheric atoms: this is the well known CRAND mechanism still invoked as one of the continuous but low intensity source of the most energetic protons populating the inner radiation belt (CRAND stands for "Cosmic Ray Albedo Neutron Decay").

It was realized later on that the Ring Current particles mainly responsible for the observed Dst geomagnetic variations have smaller energies (< 100 keV) than those produced by the CRAND low intensity sources of MeV protons populating the inner radiation belt. Spacecraft such as IMP-6, SCATHA, and AMPTE helped evolve our understanding of ring current and radiation belt particles trapped in the geomagnetic field. More recently, in the 90's, after the key observations made with the CRRES satellite during the large geomagnetic storm event of 24 March 1991 (Blake *et al.,* 1992), it has eventually become evident that, besides this slow CRAND source, coupled with diffusion in pitch angles and in radial distances, there is at least one additional source of energetic electrons and ions that accelerates and/or injects them deep into the equatorial magnetosphere, as well as in the polar cusps. This source is sporadic in nature; it is clearly controlled by the direction of the interplanetary magnetic field. Sporadic and rather catastrophic injections of relativistic electrons as well as ions with energies up to 100 MeV ions into the equatorial magnetosphere down to L = 2.5, are observed. These events are currently of great concern and primary importance in modern Space Weather projects (Blake *et al*., 1992 & 1997; Li *et al.,* 1993 & 1996; Chen *et al.,* 1997; Fritz *et al.,* 1999; Friedel *et al.,* 2002).

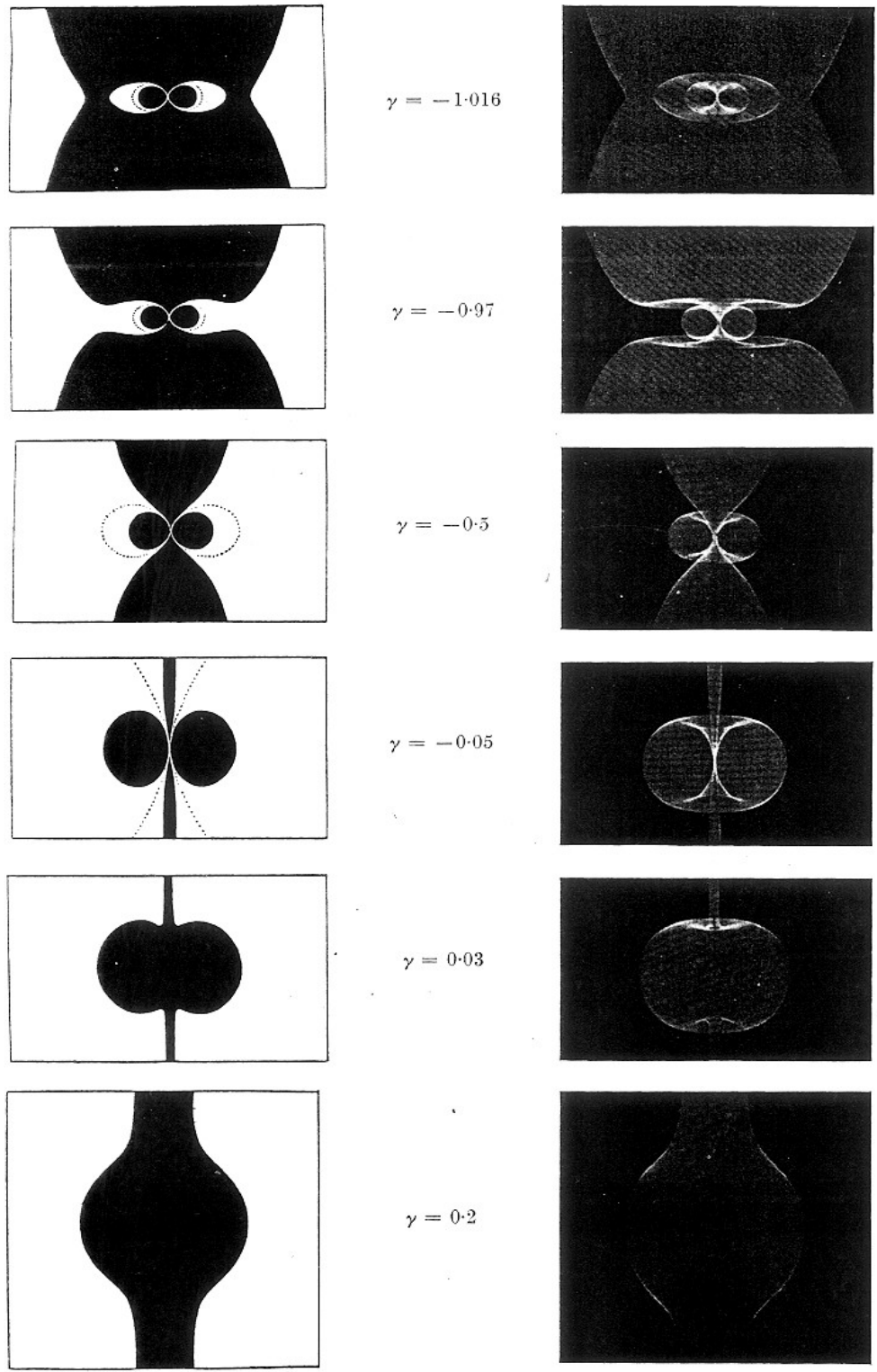

Fig. 1. *The forbidden and allowed Störmer zones for charged particle moving in a dipole magnetic field. These panels has been redrawn to be similar to the well known figure published in Störmer's book (1955).*

For historical records note also that the diamagnetic effect produced by the Ring Current on the geomagnetic cut-off latitude has been studied by Treiman (1953). The geophysical effects of an enhanced Ring Current dipole magnetic field added to the Earth dipole on the motion and fluxes of Radiation Belt particles has been addressed by Dessler and Karplus (1961) and McIlwain (1996).

An impressive amount of papers have been published based of tracing Cosmic Ray trajectories in more and more sophisticated geomagnetic field models. The emphasis in these papers was to determine the access cones, the asymptotic directions of primary Cosmic Rays and/or the cutoff rigidity for different observing stations at the Earth surface. These computer intensive trajectory simulations have enabled to investigate the effects of non-dipole components or/and time changes of the magnetic fields of the Earth, those generated by magnetospheric current systems, or the interplanetary magnetic field. These effects have been reviewed recently by Smart *et al.* (2000).

The present approach is not demanding large computer resources since it is not based on detailed trajectory tracings. In the following sections we describe an analytical approach enabling to determine the first order effects entailed by a north-southward component of the IMF on Störmer's allowed zones. For the sake of clarity, the geomagnetic field is therefore approximated by a simple dipole, ignoring all other perturbing effects due to non-dipole or non-stationary components of the geomagnetic field reviewed in the article of Smart *et al.* (2000).

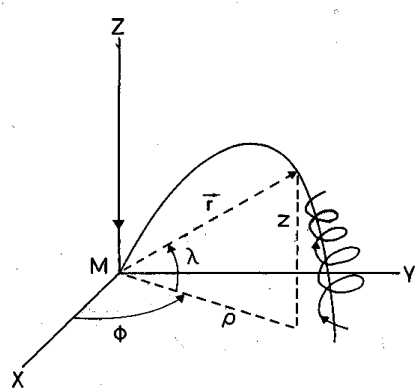

Fig. 2. *Schematic representation of a trapped charged particle motion in a magnetic dipole field.* Note that with the conventional choice of geographic coordinates ($\hat{r}, \hat{\lambda}, \hat{\varphi}$) or polar coordinates ($\hat{z}, \hat{\rho}, \hat{\varphi}$), the Earth's magnetic dipole moment *M* is anti-parallel to the $\hat{z}$-axis.

## THE MATHEMATICAL FORMULATION OF STÖRMER'S THEORY, AND ITS NEW EXTENSION

Let us consider a cylindrical coordinate system ($\hat{z}, \hat{\rho}, \hat{\varphi}$) with the origin at the center of the magnetic dipole (Figure 2). The dipole moment is anti-parallel to the $\hat{z}$-axis: $M_\rho = M_\phi = 0$; $M_z = -M$. We assume that Störmer's dipole is embedded in a uniform and time-independent Interplanetary Magnetic Field, **F,** which is either parallel or anti-parallel to the $\hat{z}$-axis. Under these circumstances **F** is independent of $\phi$, the azimuthal coordinate. For simplification we assume that it is also independent of the two other cylindrical coordinates $z$ and $\rho$. These assumptions are not critical from a physical point of view, but offer a definite advantage from the mathematical point of view; indeed, in this case the Hamiltonian (1) is still independent of the azimuthal coordinate, $\phi$.

As a matter of consequence, $p_\phi$ is again a constant of motion as in Störmer's theory. Just as in Dungey's (1961) seminal paper, **F** is either parallel ($F_z > 0$) or anti-parallel ($F_z < 0$) to the $\hat{z}$-axis. The vector potential of the combined IMF and Earth's dipole is then given by

$$A(z,\rho) = \left[ \frac{\rho F}{2} - \frac{\rho M}{r^3} \right] \hat{\varphi} \tag{2}$$

where $r^2 = \rho^2 + z^2$.

For a northward IMF, $F = F_z > 0$, while for a southward orientation $F < 0$. Although other tilt angle for **F** can be envisaged in future work, we limit the present discussion to parallel and anti-parallel orientations, since it clearly illustrates the point to be made in this first contribution. It circumscribes the effect of the IMF orientation, while keeping the formulation in a simple analytical form.

As a consequence of the axial symmetry, $p_\phi$, the azimuthal component of the generalized momentum of the particle is a constant of motion:

$$p_\varphi = m\rho^2 \frac{d\varphi}{dt} + Ze\rho A_\varphi = p_o \tag{3}$$

where $p_o$ can assume any value between $-\infty$ and $+\infty$.

The three-dimensional problem is thus reduced to the simpler two-dimensional one of finding the motion of a particle in the $(z, \rho)$ meridional plane under the influence of the potential

$$V(z,\rho) = \left[ \frac{p_o}{\rho} - ZeA_\varphi \right]^2 / 2m \tag{4}$$

The angular velocity, $d\phi/dt$, of this moving plane and of the particle can be determined by integrating Eq. (3) or

$$d\phi/dt = [p_o - Ze\rho A_\phi] / m\rho^2 \tag{5}$$

where $A_\phi$ is determined by Eq. (2), and where $\rho(t)$ and $z(t)$ must first be obtained by numerical integration of the two additional equations of motion which determine the trajectory of the particle in the meridional plane. The latter mechanical problem is similar to calculating the motion of a bead rolling in a landscape where the altitude would be proportional to the potential $V(z,\rho)$.

Dragt (1965) introduced a convenient unit of length

$$r_u = |Ze M / p_o| \tag{6}$$

This unit length differs from the Störmer unit length as indicated below (see Eq. 14); $r_u$ is inversely proportional to the angular momentum, $p_o$, while the Störmer unit depends on the modulus of momentum, $mv$, of the particles.

Furthermore, a new unit of time

$$t_u = |m(Ze M)^2 / p_o^3| \tag{7}$$

Using $A_\varphi$ from Eq. (2), the equations (1), (4) and (5) become in dimensionless form

$$H = \frac{1}{2}\left(p_z^2 + p_\rho^2\right) + V = \frac{1}{32\gamma_1^4} \tag{8}$$

$$V = \frac{1}{2}\left[\frac{\varepsilon}{\rho} - \frac{\rho}{r^3} + b\rho\right]^2 \tag{9}$$

$$\varphi(t) = sign(Z) \int \left[\frac{1}{r^3} - b - \frac{\varepsilon}{\rho^2}\right] dt \tag{10}$$

with the following definitions:

$$\varepsilon = sign\,(-p_o/Z) \tag{11}$$

$$b = \frac{F r_u^3}{2\,M} \tag{12}$$

$$\frac{1}{\gamma_1^4} = 16\, r_u^4 (mv/ZeM)^2 \tag{13}$$

where $v$ is the velocity of the particle. In dimensionless units $v = 1/4\gamma_1^2$. It can be seen that the larger the value of $\gamma_1^2$, the smaller is the velocity $v$ and the total energy $H$.

The dimensionless constant $\gamma_1$ is identical to that used by Störmer (1955) in his later work; it is equal to $-\gamma$, the parameter used in his original papers, as well as in Figure 1. The relationships between the values of $\gamma_1$ (gamma$_1$) and the kinetic energy of a proton or an electron are illustrated in Figures A1 and A2 respectively, for three values of $r_u$ (see Appendix A).

Note that in his initial work Störmer used another unit length defined by

$$C^{st} = |Ze\,M/m\,v|^{1/2} = |r_u p_o/m\,v|^{1/2} = 2 r_u \gamma_1 \tag{14}$$

In Figure A3, the Störmer unit length, $C^{st}$, is given in Earth radii ($R_E$) as a function of kinetic energy for protons (lower curve) and for electrons (upper curve). The unit of length, $r_u$, used here is not determined by the *kinetic energy* of particles, but by the value of their *angular momentum*, $p_\varphi$, which is the second constant of motion. Note that for a given value of the kinetic energy (Eq. 8), the value of $p_o$ can assume any value between $-\infty$ and $\infty$.

The time unit $t_u$ (Eq. 7) depends also on the value of $p_o$. It is plotted in Figure A4, for electrons and protons as function of the unit length $r_u$. Note that in this unit the gyro-frequency of a particle is equal to unity when the particle traverses the equatorial plane: i.e. where $\rho = r = 1$.

The dimensionless parameter, $b$, is plotted in Figure A5 versus the intensity of the interplanetary magnetic field, $F$ (in nT), for three different values of $r_u$: $r_u = 14 R_E$, $10 R_E$ & $6 R_E$.

The properties of the Störmer potential (Eq. 9) are outlined in the following sections for $\varepsilon = \pm 1$: (i) for a pure dipole as considered in Störmer's work; (ii) for a northward IMF, and (iii) for a southward IMF. The distribution of magnetic field lines in the meridian plane is illustrated in Figure 3 for all these three different cases.

# THE STÖRMER CASE ($F = 0$)

When the interplanetary magnetic field intensity $F$ is set equal to zero, $b$ is equal to zero in Eqs. 9, 10 and 12; the original expressions of Störmer's theory are then recovered. The magnetic field lines are all 'closed' as illustrated in Figure 3a.

### The Störmer potential

A 3-D representation of the Störmer potential (9) is illustrated in Figure 5a for $\varepsilon = +1$. The lines separating the different colors/shading correspond to equipotential isocontours of $V(z,\rho)$. The top panel in Figure 5a is a projection showing a 2-D map of these isocontours which is equivalent to the line contour plots of Figure 4. The dashed lines correspond to a series of values of $\gamma_1$ and decreasing or increasing values of $V(z,\rho)$. They are equivalent graphical representations of the boundaries separating Störmer's allowed and forbidden zones already presented in Figure 1, but in a different format for different values of the constant of motion $\gamma$.

The deep valley near the origin in the 3-D "landscape" and 2-D maps contains the Thalweg. The "Thalweg" corresponds to the magnetic field line whose equatorial distance is equal to $\rho = 1$, in dimensionless variables; the equation of this field line is given by

$$r = cos^2(\lambda) = \rho^2 / r^2 \qquad (15)$$

The "Thalweg", was shown by the dotted line in the upper panels of Figure 1 for which $\gamma < 0$; in Figure 4, it corresponds to the dotted line crossing the equatorial plane $z = 0$, at $\rho = 1$. Along the Thalweg $V = 0$. On both sides of this deep "valley", the potential $V(z,\rho)$ increases steeply, as illustrated in Figures 4 and 5a.

According to Eq. (8), any particle that has a given total energy $H_o < 1/32$, is characterized by $\gamma_1 > 1$; the kinetic energy associated with the motion in the co-moving meridian plane tends to zero when the particle approaches the isocontour $V = H_o$. Note that $v_\rho^2 + v_z^2 = 0$ and $v_\phi = v$ at this location. A particle for which $\gamma_1 > 1$ belongs to the class $\gamma < -1$ in Figure 1; it can be either permanently trapped in the valley if $\rho < 2$, like Van Allen particles, or it will be deflected away from the Earth downhill the landscape beyond the outermost isocontour $V = H_o$, if this energetic particle comes from the Sun or from elsewhere in the solar system.

The dashed lines in Figure 4 running along the innermost "wall" close to the Earth determine the smallest distance of penetration for interplanetary particles: i.e. the "geomagnetic cut-off" for energetic interplanetary particles of a given energy $H$ (i.e. a given value of $\gamma_1$). Any particle that has a total energy $H$ smaller than $1/32$ (i.e. for wich $\gamma_1 > 1$) cannot penetrate from outside into the valley, nor can it escape out of it when it is inside the inner allowed Störmer region See in a following section a more comprehensive discussion of the geomagnetic cut-off and of its dependence on the kinetic energy and azimuthal component of the generalized momentum of the particles, as well as on the intensity of the southward interplanetary magnetic field.

Note that the plots 4 and 5a correspond to cases for which the azimuthal component of the generalized momentum of the particle is such that $p_\phi / Z$ is negative (i.e. $\varepsilon = +1$, according to Eq. 11). Only interplanetary particles for which $\varepsilon = +1$ have access to the Earth surface, provided that $0 < \gamma_1^4 < 1$ (or $-1 < \gamma < 0$); otherwise, when $\gamma_1^4 > 1$ (∴ $H_o < 1/32$), the access "pass" in the equatorial region is closed, indeed then, the particles have not enough energy to run over this "pass". This "pass" is a saddle point in the "landscape".

Note that there is no "valley", nor "Thalweg" when $\varepsilon = -1$, since in this case there are no points where $V = 0$, except at infinity. This unfavorable case is illustrated in Figure 5b. This situation corresponds either (i) to electrons ($Z = -1$) which have an azimuthal component of the generalized momentum directed westward ($p_\phi = p_o < 0$), or (ii) to positively charged ions ($Z > 0$) which have an azimuthal component of the generalized momentum directed eastward ($p_\phi = p_o > 0$). All those particles for which $\varepsilon = -1$ are on orbits that belong to the class $\gamma > 0$ in Figure 1. They can never reach the Earth's surface, since they are all deflected away from the geomagnetic dipole. Therefore these interplanetary particles are of no special interest for geophysical applications, since they have no access to the magnetosphere nor to the terrestrial atmosphere.

Let us go back to particles for which $\varepsilon = +1$. On the opposite side of the Thalweg valley there are other steep "walls" deflecting particles towards the equatorial plane. At $\rho = 2$, in the equatorial plane, there is a saddle point (see Figures 4 and 5a). Particles located at this saddle point and for which $v_\rho^2 + v_z^2 = 0$, have a Larmor gyroradius which is equal to the radial distance: $\rho = 2$ or $\rho = 2\, r_u$ in dimensional units. This circular trajectory encircling the Earth in the equatorial plane is a special periodic orbit. It has been comprehensively studied by Graef and Kusaka (1938) and Avrett (1962). There are several other periodic orbits and quasi-periodic ones that have been studied extensively by students and colleagues of Lemaître, at the Université catholique de Louvain.

The saddle point in Figures 4 and 5a is a mathematical singularity where the downhill slopes of the "landscape" point toward the equatorial plane, except in the vicinity of the equatorial plane where they are directed away from the saddle point. The direction of the slope is indicated by the direction of the arrows in Figure 4. The value of $V$ is equal to $V_P = 1/32$ at the saddle point, as well as all along the dashed isocontour labeled $V = 1/32$.

Figure 6 shows the equatorial cross-section of the surface $V(z,\rho)$ for different values of the IMF intensity, and for $\varepsilon = +1$. The solid line corresponds to the case of Störmer's dipole: $F = 0$. It has a minimum ($V = 0$) at $\rho = 1$ coresponding to the Thalweg; it has a maximum ($V = 1/32$) at $\rho = 2$, the position of the saddle point; beyond this point $V$ tends asymptotically to zero when $\rho$ increases to infinity.

The other curves in Figure 6 correspond to non-zero values of the IMF: the dashed lines are for a southward orientation of the Interplanetary Magnetic Field ($b < 0$), the dotted lines are for a northward IMF ($b > 0$). These cases will be discussed in the following sections.

### Magnetic field line distributions

The distributions of magnetic field lines corresponding to different values of the IMF intensity are illustrated in Figures 3b-d. The dimensionless equation of magnetic field lines is

$$r = k\, cos^2(\lambda)\, [1 - b\, r^3] \qquad (16)$$

where $k$ is a dimensionless parameter generating the whole distribution of magnetic field lines when $k$ varies from 0 to infinity.

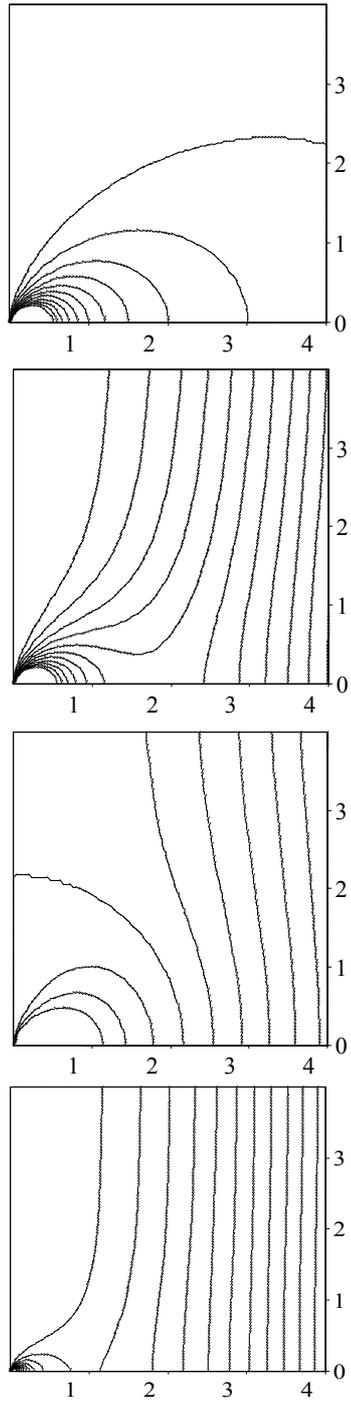

Fig. 3. *Distribution of magnetic field lines* for different values
and orientations of the interplanetary magnetic field :
  *a)* upper left : Störmer case : $b = 0$;
  *b)* upper right : southward IMF : $b = -0.1$;
  *c)* lower left : northward IMF : $b = 0.1$;
  *d)* lower right : southward IMF : $b = -0.6$.

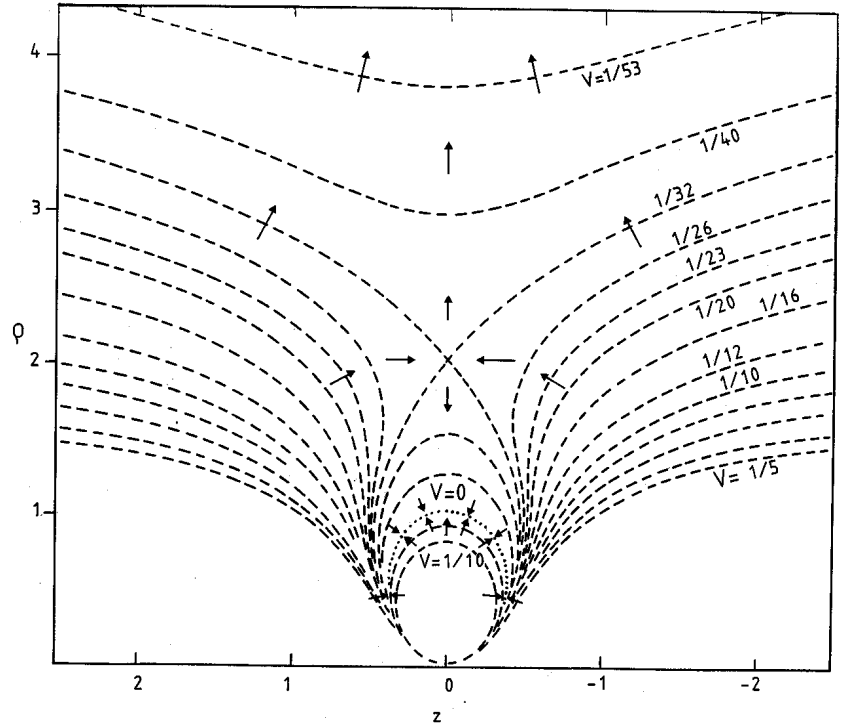

Fig. 4. *Isocontours of the Störmer dimensionless potential, V(z,ρ)* [eq. (8) for $F = b = 0$, and $\varepsilon = sign(-p_\phi/Z) = +1$]. The Thalweg shown by the dotted line is the dipole magnetic field line ($r = cos^2\lambda$) around which a low energy particle spirals, and along which its guiding center oscillates between two conjugate mirror points. The dashed lines determine the inner and outer boundaries of the allowed and forbidden Störmer zone for a series of values of the potential energy (*V*) ranging from $V = 0$ (corresponding to the Thalweg) to $V = 1/53$. The arrows indicate the direction of the gradient of *V(z,ρ)*. Note that $V = 1/32$ is a critical value for which the inner and outer allowed zones connect each other at the saddle point ($z = 0, \rho = 2$). This corresponds in fig. 1 to the case for which $\gamma = -1$. Indeed, when the total energy of a particle is larger than $H = 1/32$, its kinetic energy is large enough to overcome the magnetic potential barrier and penetrate deep into the geomagnetic field.

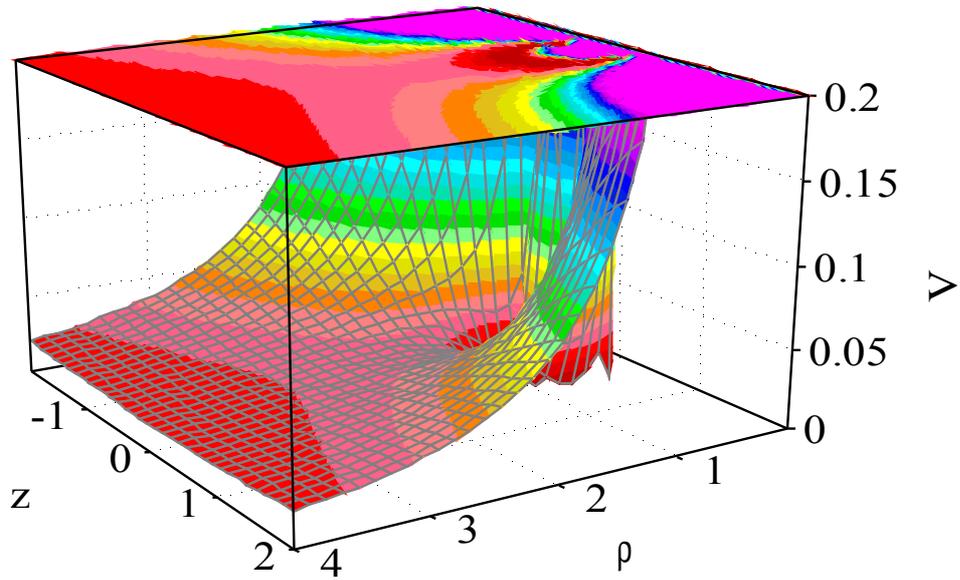

Fig. 5a. *A 3-dimensional representation of the dimensionless Störmer potential* [eq. (8) for $F = b = 0$, and $\varepsilon = sign(-p_o/Z) = +1$] the distances are in $r_u$ units given by eq. (6). The different colors/shadings separate equipotential contours. The deep valley contains the Thalweg where $V = 0$.

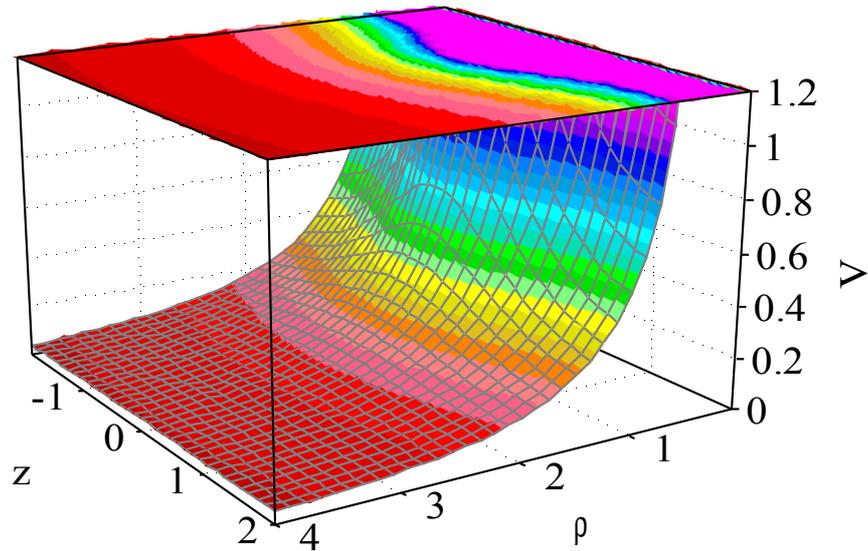

Fig. 5b. *A 3-dimensional representation of the dimensionless Störmer potential* [eq. (8) for $F = b = 0$, and $\varepsilon = sign(-p_o/Z) = -1$]. The distances are in $r_u$ units given by eq. (6). The different colors separate equipotential contours.

When $b = 0$ and $k = 1$, Eq.(15) for the Thalweg in a dipole magnetic field is recovered. When $b \geq 0$ geomagnetic field lines cross the equatorial plane ($\lambda = 0$) for all values of $k$: i.e. all magnetic field lines traversing the Earth surface are "closed", except those two emerging from the magnetic poles (see Figure 3c). However, when $b < 0$ magnetic field lines are closed only if they traverse the surface of the Earth below a critical latitude $\lambda_E$. All those penetrating this surface at higher latitudes are so called "open" (whatever that means for magnetic field lines!) or "interconnected"[♥] with the interplanetary magnetic field lines (see Figures 3b & 3d).

Magnetic field lines passing through a so-called "X-line" or "neutral line" where $B = B_z = 0$ in the equatorial plane traverse the Earth surface at the latitude $\lambda_x$ determined by

$$cos^2(\lambda_X) = r_E / k_X (1 - b\, r_E^3) \qquad (17)$$

where $k_X = 2/[3\,(-2\,b)^{1/3}]$, and where $r_E = R_E / r_u$ is the dimensionless Earth radius using the unit length of Eq. (6).

It can be verified that when $k$ is smaller than the limiting value, $k_o = r_E /(1 - b\, r_E^3)$, all magnetic field lines penetrating the Earth's surface are "open". For $k > k_o$ polar cap magnetic field lines are interconnected or "open" beyond the latitude $\lambda_X$.

Magnetic field lines corresponding to the Thalweg are always closed in Störmer's case. However, they can become open when $b < b_T = -4/(3)^3 = -0.148148$. From Figure 5 it can be seen that this threshold corresponds to $F = 42.75\ nT,\ 9.2\ nT$, and $3.4\ nT$, respectively for $r_u = 6R_E,\ 10R_E$ and $14R_E$.

The equatorial distance where the last closed Thalweg crosses the equatorial plane is at $\rho = 1.5$. The latitude $\lambda_T$ where the Thalweg magnetic field line traverses the Earth surface is given by an expression similar to Eq. (17), but for $k = 1$ and $b = -0.148148$. Note that $\lambda_T < \lambda_X$ when $b > b_T$; while $\lambda_T > \lambda_X$ when $b < b_T$.

The value of $\lambda_X$ decreases, when the southward component of the IMF intensity increases (becomes more negative), while the value of $\lambda_T$ does not change significantly with $b$, unless $-b r^3_E$ assumes extrem values of the order of 1. Note however, that the value of $\lambda_T$ is a sensitive function of the azimuthal component of the generalized momentum through definition (6) of $r_u$.

---

[♥] "Interconnected magnetic field lines" should not be confused with so called "reconnected magnetic field lines", since the latter implicitly implies that there is a non-zero electric field perpendicular to the direction of the magnetic field lines at any point where $B=0$. This "reconnection or merging electric field" accelerates particles and is often said to "convert magnetic energy into kinetic energy by accelerating the plasma particles within an elongated region". The X-line is a place where the convection electric velocity $\mathbf{E} \times \mathbf{B}/ B^2$ diverges and assumes infinitely large values. This is where ideal MHD breaks down. However, in the case of "interconnected magnetic field lines" the perpendicular electric field - when there is any - is not assumed to be *non-zero* where the B-field intensity goes to zero. Indeed, both $\mathbf{E}$ and $\mathbf{B}$ vectors can change directions or vanish at the same points; in these circumstances the convection velocity does not diverge anywhere along any magnetic field line; at X-lines like those existing magnetic distributions illustrated in Figures 3b & 3c. The convection velocity $\mathbf{v}$ remains finite, and $\mathbf{E} = \mathbf{v} \times \mathbf{B}$ is equal to zero where $\mathbf{B} = 0$.

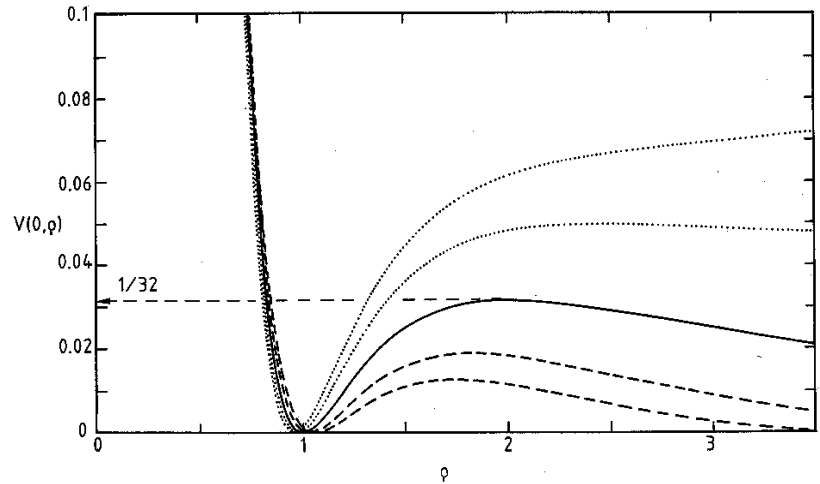

Fig. 6. Equatorial cross-section of the generalized Störmer potential (V) as a function of the dimensionless equatorial distance (ρ). When the value of the IMF is equal to zero (F = b = 0), the maximum value of $V(0, \rho_{max}) = 1/32 = 0.03125$; it is located at $\rho_{max} = 2$. For $F \neq 0$ the value of this maximum value and its position vary as described in the text. The two lower curves lines correspond to $b = -0.03$ and $-0.05$ respectively; the two upper ones to $b = +0.03$ and $+0.05$ respectively.

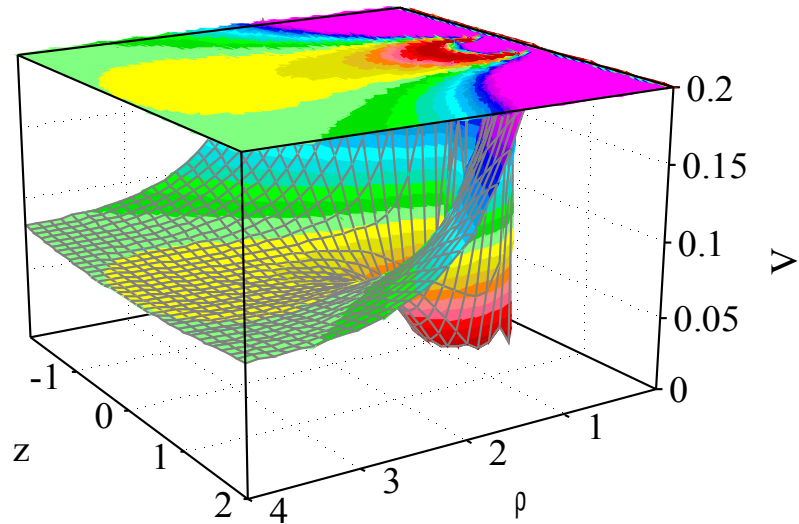

Fig. 7. *A 3-dimensional representation of the dimensionless Störmer potential* [eq. 2] *for Northward IMF : $b = +0.05$, and $\varepsilon = sign(-p_o/Z) = +1$*]. The different colors/shadings separate equipotential contours. The deep valley contains the Thalweg where $V = 0$.

## NORTHWARD IMF ($F > 0$)

When the Interplanetary Magnetic Field is Northward, $b$ is positive, and the distribution of magnetic field lines is illustrated in Figure 3b. In this case there are two points along the polar axis where the magnetic field intensity is equal to zero. These neutral points are located at a radial distance $z = \pm 1 / (b)^{1/3}$.

No geomagnetic field line is then interconnected with those of interplanetary space, except those two from the north and south poles. All field lines are draped around a closed surface: the magnetopause, corresponding to the envelop of early days closed magnetospheric models. It is evident that this particular orientation of the IMF is the most unfavorable for energetic particles as well as low energy solar wind particles to access the inner regions of the geomagnetic field. The area of the polar cap regions that is accessible to these particles via interconnected magnetic field lines, increases gradually when the tilt angle of the IMF increases from zero ($b > 0$) to 180° ($b < 0$).

The Thalweg is located at smaller radial distances when $b > 0$ than in the case of a pure dipole (see dotted curves in Figure 6). Note, however that the latitude, $\lambda_T$, where the Thalweg magnetic field lines intersects the Earth surface is not very sensitive to the value of $b$, but instead depends more importantly on the value of the azimuthal component of the generalized momentum $p_o$, as indicated above.

Increasing the northward component of the IMF pushes the saddle point further away from Earth, to radial distances larger than $\rho = 2$. Furthermore, the "altitude" of this "pass" becomes larger than the critical value of $V_P = 1/32$ corresponding to Störmer's case (solid curve in Figure 6). This implies that only interplanetary particles with larger total energies (i.e. values of $\gamma_l$ smaller than $1$) will be able to overcome the magnetic potential barrier and invade the Thalweg valley. Another practical consequence is that northward turning of the IMF increases the energy threshold for which trapped Van Allen particles can escape out of the modified inner allowed Störmer zone. This implies that when the IMF turns northward, particles of a fixed energy tend to become more tightly trapped within the valley forming the inner allowed region♠.

A 3-dimensional representation of the dimensionless potential, $V$, is shown in Figure 7, for $b = +0.05$ and for $\varepsilon = +1$. At large radial distances the total magnetic field intensity becomes uniform and equal to $F$; the Störmer potential (Eq. 9) increases indefinitely as $½(b\rho)^2$ when $\rho \to \infty$. This asymptotic variation at large radial distance is the same whether the IMF is northward or southward, and whether $-p_\phi / Z$ or $\varepsilon$ are positive or negative. This trend is not apparent in Figures 6 and 8 since for $|b| < 0.05$ the positive slope develops only beyond $\rho = 4$, but it is clearly seen in Figure 9 for $b = -0.3$.

## SOUTHWARD IMF ($F < 0$)

When the Interplanetary Magnetic Field is southward, $b$ is negative. The field lines corresponding to $b < 0$ are illustrated in Figures 3b & 3d for $b = -0.1$ & $-0.6$.

**Magnetic field line distribution**

There is a field line where the magnetic field intensity is equal to zero (X-line), as in the magnetic field model employed by Dungey to introduce the concept of reconnection or merging; here, this X-line or neutral line is a circle located in the equatorial plane at a radial distance $\rho_x = 1 / (-2b)^{1/3}$. To shift the equatorial radius of the X-line down to $\rho_x < 1$, the value of $-b$ must be larger than $0.5$. According to Figure A5 this requires rather extrem values of the southward IMF component.

All geomagnetic field lines passing through the X-line, as well as those emerging from the polar caps at latitudes larger than $\lambda_X$, are open or interconnected with those of the interplanetary magnetic field. This latitude that corresponds to the first interconnected geomagnetic field line, decreases when $-b$ increases. The larger the intensity of the southward component of the IMF, the smaller is $\lambda_X$ and the larger is the polar cap area with geomagnetic field lines that are interconnected with the IMF.

---

♠ Note that this effect (enhanced trapping efficiency) applies whether Alfvén conditions are satisfied or not: i.e. whether or not the first order guiding center approximation is a satisfactory approximation describing Störmer's exact orbital motion.

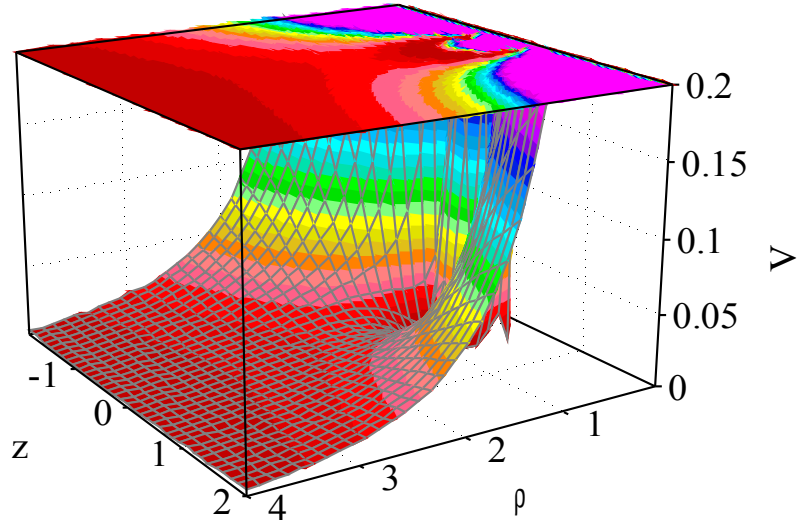

Fig. 8. A *3-dimensional representation of the dimensionless Störmer potential* [eq. 8] for Southward IMF : *b = -0.03* and *ε = sign(-p₀/Z) = +1*. The different colors/shadings separate equipotential contours. The deep valley contains the Thalweg where *V = 0*.

**The modified Störmer potential**

From the dashed lines in Figure 6 it can be seen that the Thalweg is pushed to larger radial distances when $-b$ increases, while the saddle point is pulled closer to Earth. Furthermore, when *b* becomes more negative, the altitude of this "pass" decreases below the value $V_P = 1/32$, corresponding to the Störmer case.

There is a critical value $b_T = -4/(3)^3 = -0.148148$ for which the equatorial distances of the Thalweg and of the saddle point coincide; they are then both collocated at $\rho = 1.5$. For *b = -0.148148* the saddle point disappears, and its "altitude" is reduced to zero : *V(0, ρ =1.5) = 0*. From Figure A5 it can be seen that this critical value of *b* corresponds to a southward interplanetary magnetic field intensity ranging between *-3.4nT* and *-42.75nT*, for $r_u$ ranging between *14$R_E$* and *6$R_E$*, respectively.

For even larger southward IMF values (*b < -0.148148*) a new saddle point pops up at *ρ < 1.5*. The Störmer potential transforms then into the new landscape illustrated in Figure 9. The function *V(z,ρ)* still has maximas at *ρ = ∞* and *ρ = 0*, but it has a minimum at a new saddle point where the vectors *-grad V* are now directed away from this saddle point in the directions perpendicular to the equatorial plane, and toward this singular point in the directions parallel to the equatorial plane: this is in opposite directions to the arrows shown around the former saddle point in Figure 4.

When *b < -0.148148* the Thalweg along which *V = 0* is no longer a closed field line as in Störmer's case, but it is formed of two open magnetic field lines rooted in the polar caps at latitude $\lambda_T$. These interconnected geomagnetic field lines extend to infinity like all other interconnected magnetic field lines illustrated in Figures 3b and 3d.

The equipotential lines of the modified Störmer function (9) are then illustrated in Figure 10, using a similar format as in Figure 4. Note that the Thalweg (not plotted in Figure 10) is located in the cusp like valley colored in dark-red, and extends to infinity *z = ±∞* in the northern and southern hemispheres.

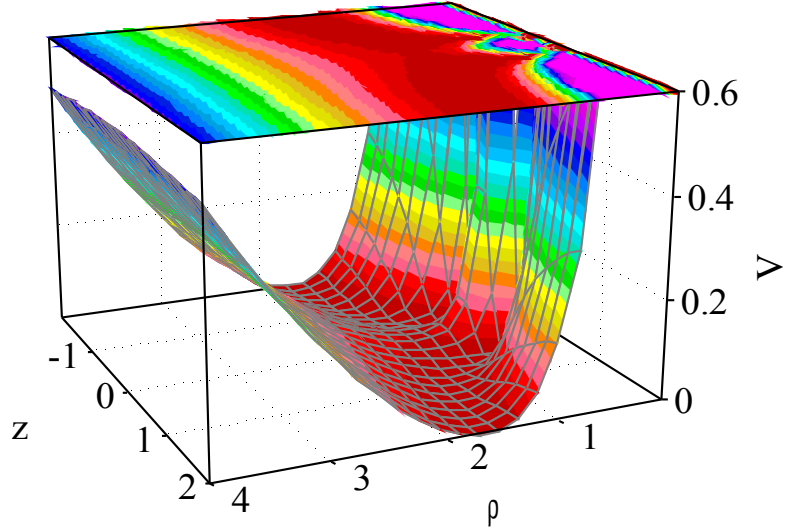

Fig. 9. *A 3-dimensional representation of the dimensionless Störmer potential* [eq. 8] for a large Southward IMF : $b = -0.3$ and $\varepsilon = sign(-p_o/Z) = +1$. The different colors/shadings are separated by equipotential contours. The Thalweg where $V = 0$ is now formed of two separate branches linking the high latitude cusp regions on both hemispheres and interplanetary medium. Particles coming from infinity along these field lines can penetrate down to the Earth surface without having to bridge over any magnetic potential barrier as in the case when $b > -0.148148$, as for instance in the Störmer theory when $b = 0$.

Evidently, interplanetary particles of all energies coming from $z = \pm\infty$ can then freely spiral into the polar cusps along the Thalweg, as well as along all other interplanetary magnetic field lines which are interconnected with those of the Earth dipole. Of course, those spiraling along the Thalweg branches do not have to overcome any magnetic potential barrier since $V = 0$ all along these special field line. This is not the case, however, for particles spiraling into the magnetosphere along all other interconnected magnetic field line located at smaller or larger latitudes. Indeed, along these other field lines the modified Störmer potential has a non-zero maximum value that determines the energy threshold of the particles that may rain into the magnetosphere over the polar caps.

**The equatorial distances and latitudes of the geomagnetic cut-off**

In Figures 10 and 4, the intersection of the different equipotential lines with the Earth's surface ($r = r_E$ ; not plotted here) determines the latitude of the so-called "geomagnetic cut-off" for charged particles of a fixed energy $H = 1/32\gamma_1^4$, and for a fixed azimuthal component of the generalized momentum $p_o$. It is the locus of points closest to Earth surface, where $v_\rho^2 + v_z^2 = 0$ and $v_\phi = v$. The equatorial distance of the geomagnetic cut-off, $\rho_G$, is determined by the smallest positive root of the algebraic equation:

$$\rho_G^2 / (1 - \rho_G - b\,\rho_G^3) = 4\,\gamma_1^2 \qquad (18)$$

Since $|b\rho_G^3| \ll 1$, the value of $\rho_G$ is not a very sensitive function of the value interplanetary magnetic field intensity. It depends of course significantly on the value of $\gamma_1^2$ : i.e. on the energy $H$ of the particle. The larger this energy, the smaller is $\gamma_1$, and the smaller is $\rho_G$. In other words, the larger the kinetic energy of the energetic interplanetary particles, the deeper they can penetrate into the equatorial plane of the magnetosphere. This is precisely what is observed during a geomagnetic storm when energetic particles are injected into the equatorial region of the magnetosphere during a period of prolonged southward IMF (McIlwain, 1966; Baker *et al.*, 1992, 1997).

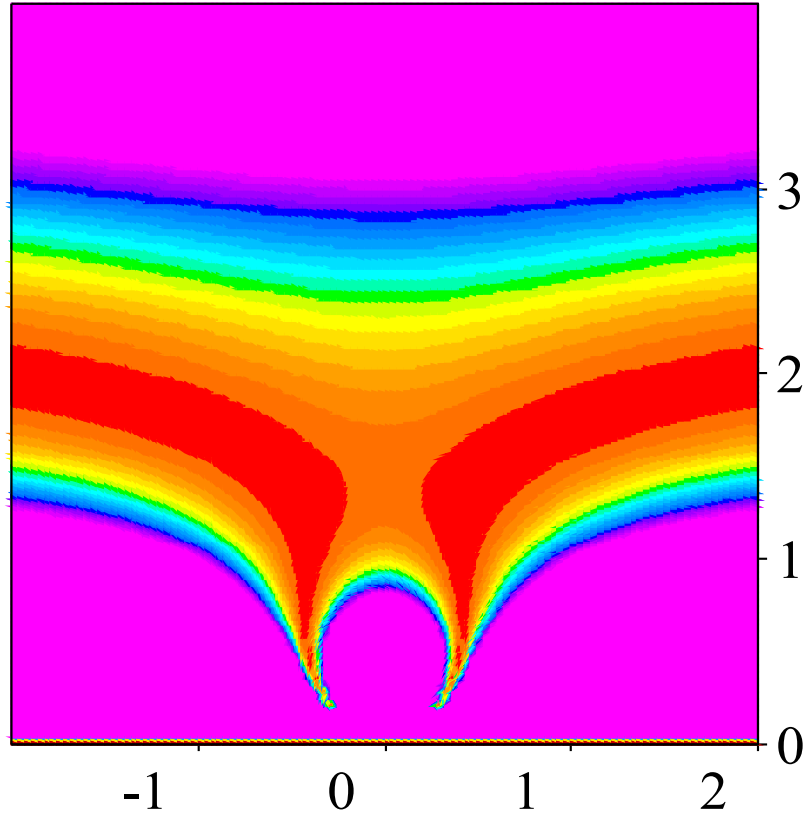

Fig. 10. *Isocontours of the Störmer dimensionless potential, $V(z,\rho)$* [eq. (8)] for a relatively large (negative) southward IMF : $b = -0.21$, ( i.e.: $F = -13.4$ nT if $r_u = 10\ R_E$) and for $\varepsilon = sign(-p_o/Z) = +1$. The new saddle point is at equatorial distance $\rho = 1/(-2\ b)^{1/3} = 1.33$ in units $r_u$. The corresponding value of the Interplanetary magnetic field (in nT) can be determined from fig. A5 for different values of $r_u$ (in Earth radii). The Thalweg extends now to infinity and is located along two interconnected magnetic field lines in the middle of the dark-brown through-like regions in both hemispheres. Along this open Thalweg particles of any energy can spiral from infinity ($z = \pm\infty$) to the Earth surface along cusp-like interconnected magnetic field lines.

The geomagnetic cut-off determines also the latitude, $\lambda_G$, above which primary cosmic rays particles of dimensionless energy $H > 1/32\gamma_I^4$ and azimuthal component of the generalized angular moment $p_o$ can be observed at the surface of Earth. This latitude is the smallest solution of the equation :

$$cos^2(\lambda_G)\ (1 - b\ r_E^3) + r_E^2\ cos(\lambda_G)/4\ \gamma_I^2 - r_E = 0 \qquad (19)$$

To a good approximation $\lambda_G = arcos(r_E^{1/2})$, since $r_E^2 / 4\ \gamma_I^2 \ll 1$. The value of $\lambda_G$ is not very sensitive of the interplanetary magnetic field intensity, but it depends strongly on the value of $r_E$. The larger the value of the angular momentum, $p_o$, the smaller the value of $r_u$ (Eq. 6), the larger $r_E^{1/2}$, and the smaller is $\lambda_G$.

## DISCUSSION AND CONCLUSIONS

When the IMF turns southward, a larger fraction of polar cap geomagnetic field lines become interconnected with the interplanetary magnetic field lines, at latitudes larger than $\lambda_X$ given by Eq. (17). This latitude is determined by the intensity of the southward component of the IMF, and decreases when $-F$ or $-b$ increase.

Furthermore, when the IMF turns southward, the topology and shapes of Störmer's allowed and forbidden regions for energetic charged particles are also modified in an interesting and most relevant manner. A southward turning of the interplanetary magnetic field allows an easier access of energetic electrons and ions into the inner magnetosphere, by lowering the magnetic potential barrier located at the saddle point in the Störmer potential. The "pass" between the inner and outer allowed zones opens up, when $-F$ increases.

This allows particles of lower energies trapped in the Thalweg valley to escape out of it more easily than in the case considered by Störmer for $b = 0$. It also allows interplanetary particles with lower kinetic energy to flow into the inner magnetosphere. These particles may penetrate as deep as the lower edge of the innermost allowed region. This inner edge determines the geomagnetic cut-off whose equatorial distance is given by Eq. (18), and the latitude by Eq. (19).

The isocontours in Figures 4 and 10 don't correspond to magnetic field lines, but they determine the geomagnetic cut-off where particles of a given energy and azimuthal component of the generalized momentumare deflected: i.e. where their velocity component parallel to the co-moving meridian plane vanishes: $v_\rho^2 + v_z^2 = 0$. When a particle reaches the geomagnetic cut-off surface its velocity becomes perpendicular to this meridian plane: $v = v_\phi$.

On the other hand when the IMF turns northward, only two geomagnetic field lines are interconnected to the northward IMF. These are the field lines emerging from the north and south magnetic poles.

When the value of $F$ is positive and increases, the "pass" between the inner and outer allowed zones shrinks, and the magnetic potential barrier between the allowed Störmer zones levels up. This makes it more difficult for the particles trapped in the Thalweg valley to escape out of the inner allowed region. It also makes it more difficult for interplanetary particles of a given kinetic energy and azimuthal component of the generalized momentumto penetrate into the magnetosphere.

Of course, the presence of magnetopause currents as well as of the turbulent magnetic fields in the magnetosheath complicates the topology of the actual magnetic field distribution, and leaves no hope to keep the $\phi$-invariance which enabled us here to treat this problem analytically. Nevertheless, we expect that the conclusions outlined above remain applicable to more sophisticated models for the geomagnetic field taking into account small perturbing internal and external current systems and for which the dipole component is the dominant one.

This theoretical study definitely confirms, without detailed and CPU time demanding tracings of charged particles in high order approximations of the geomagnetic field, that the IMF has a definite influence on the access of energetic interplanetary particles into the magnetosphere.

Beside recalling and updating the seminal work of Störmer at the beginning of the 20$^{th}$ century, this study indicates how this theory can be extended to take into account the major effects of an external interplanetary magnetic field. But a detailed discussion of the geophysical consequences of this generalization of Störmer's theory are beyond the scope of this first paper, and should be examined in forthcoming studies.

# APPENDIX A

# CONSTANTS OF MOTION, LENGTH UNIT AND TIME UNIT

Using the length unit and time unit defined by Eqs. (6) and (7), the total energy is given by Eq. (8), where $\gamma_1$ (gamma$_1$) is plotted in Figure A1, versus kinetic energy of protons (in Mev), for three different values of the unit length $r_u$. The value of $\gamma_1$ for electrons with energies ranging between 100 keV and 1 MeV is shown in Figure A2 for the same set of values of $r_u$.

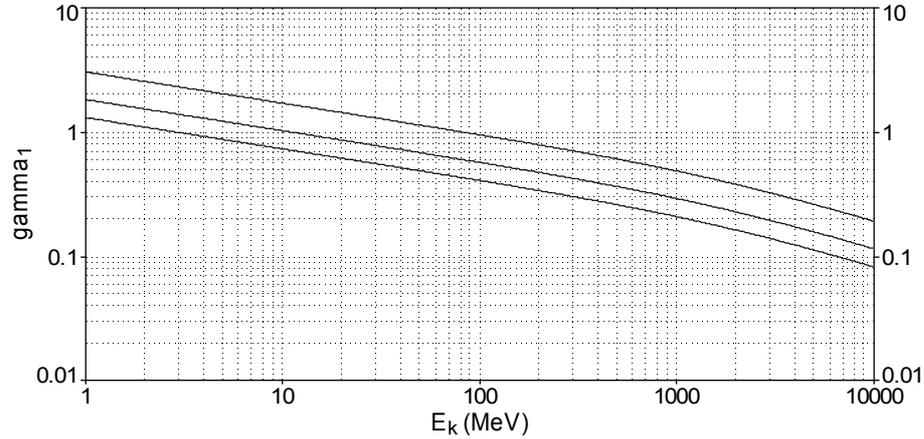

Fig. A1. *Dimensionless constant of motion $\gamma_1$ versus proton kinetic energy (in MeV)* [eq. (13)] for different values of the unit of length: $r_u = 6R_E$ (upper curve); $r_u = 10R_E$ (second curve); $r_u = 14R_E$ (lower curve).

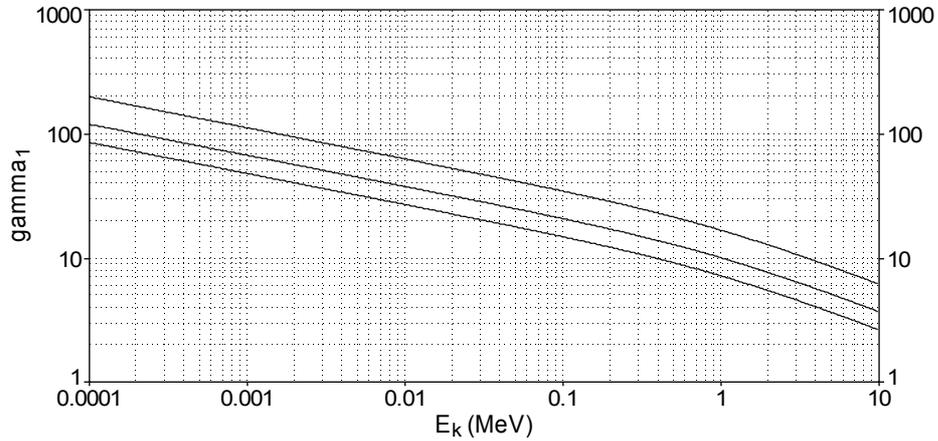

Fig. A2. *Dimensionless constant of motion $\gamma_1$ versus electron kinetic energy (in MeV)* [eq. (13)] for different values of the unit of length: $r_u = 6\ R_E$ (upper curve); $r_u = 10\ R_E$ (second curve); $r_u = 14\ R_E$ (lower curve).

In his initial studies of the motion of charged particles in a magnetic dipole Störmer used a different length unit defined by Eq. (15). This length unit, $C^{st}$, called "the Störmer", is determined by the momentum of the particles. Figure A3 shows the values of $C^{st}$ (in Earth's radii) for protons and electrons versus their kinetic energy (in MeV).

The unit of time $t_u$ used in this article is defined by Eq. (7). It is plotted in Figure A4 as a function of $r_u$ (in $R_E$) for protons and electrons.

The basic difference between the theory outlined in the present article and that of Störmer resides in the addition of a northward of southward Interplanetary Magnetic Field : $F$. The relationship between $F$ and the dimensionless parameter $b$ defined by Eq.(12) is plotted in Figure A5 for three different values of the unit length $r_u$.

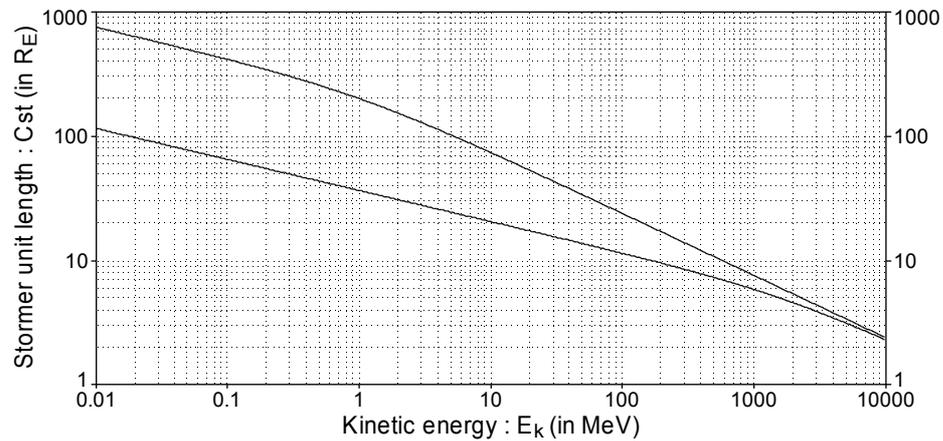

Fig. A3. *Störmer unit length (Cst)* in Earth's radii [eq. (15)] as a function of the kinetic energy of protons (lower curve) and of electrons (upper curve)

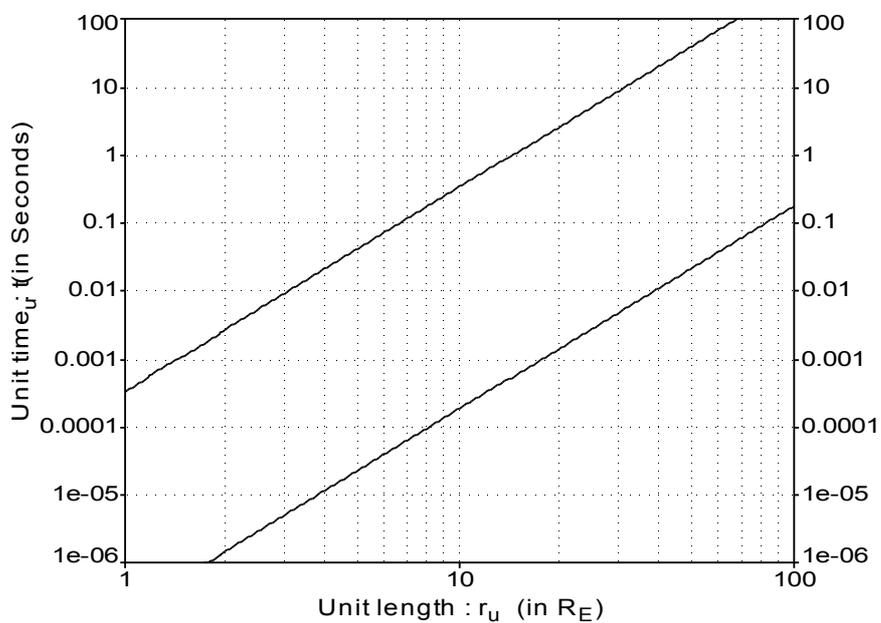

Fig. A4. *Unit length $t_u$* in seconds [eq. (7)] as a function of the unit length $r_u$ for protons (upper curve) and electrons (lower curve)

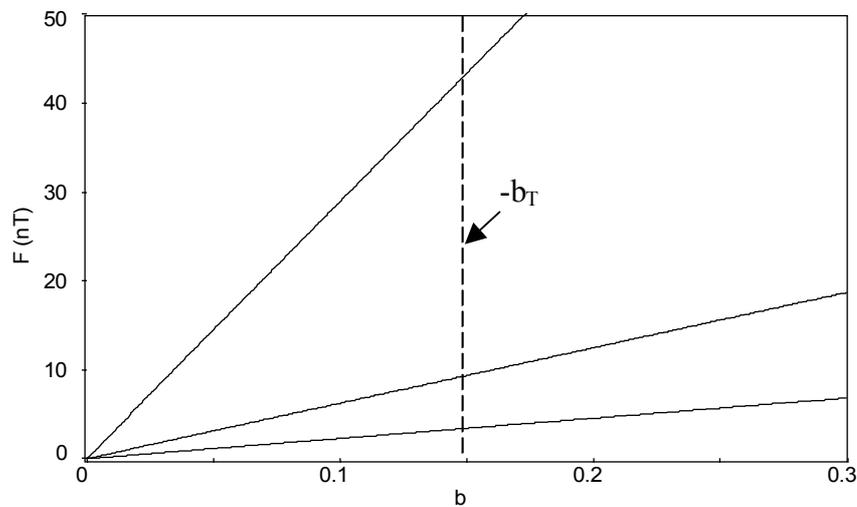

Fig. A5.  *Interplanetary Magnetic Field intensity (in nT) versus the modulus of the dimensionless parameter b* [eq. (12)] *for different values of the unit of length:  $r_u = 6\ R_E$ (upper curve); $r_u = 10\ R_E$ (middle curve);   $r_u = 14 R_E$ (lower curve).*

# ACKNOWLEDGMENTS

I wish to thank F.S. Singer (SEPP), D. Heynderickx (BIRA-SPENVIS), V. Pierrard (IASB), H. Lamy (IASB), and V. Cadez (BIRA) for useful remarks and comments. This work was undertaken at IASB-BIRA, Brussels. It was funded by the OSTC (*Federal Office for Scientific, Technical and Cultural Affairs* ; http://www.belspo.be/belspo/ ) under the PRODEX-CLUSTER Contract 13127/98/NL/VJ/(IC).